\documentclass{pasj00}
\begin{document}

\SetRunningHead{A. Bamba et al.}{Transient X-Ray Pulsar AX J1841.0$-$0536}
\Received{2001 June 19}%{yyyy/mm/dd}
\Accepted{2001 September 12}%{yyyy/mm/dd}

\title{Discovery of a Transient X-Ray Pulsar, AX~J1841.0$-$0536,
in the Scutum Arm Region with ASCA}

%%% begin:list of authors
\author{Aya \textsc{Bamba}, Jun \textsc{Yokogawa},
Masaru \textsc{Ueno}, and Katsuji \textsc{Koyama}}
\affil{Department of Physics, Graduate School of Science, Kyoto University, 
Sakyo-ku, Kyoto\ 606-8502}
\email{bamba@cr.scphys.kyoto-u.ac.jp,
jun@cr.scphys.kyoto-u.ac.jp,
masaru@cr.scphys.kyoto-u.ac.jp,
koyama@cr.scphys.kyoto-u.ac.jp}
\and
\author{Shigeo \textsc{Yamauchi}}
\affil{Faculty of Humanities and Social Sciences, Iwate University, 
3-18-34 Ueda, Morioka, Iwate 020-8550}
\email{yamauchi@hiryu.hss.iwate-u.ac.jp}
%%% end:list of authors

\KeyWords{stars: neutron --- stars: pulsars: individual (AX~J1841.0$-$0536)
--- X-rays: spectra --- X-rays: stars --- X-rays: transient}

\maketitle

\begin{abstract}

We report on the discovery of a transient X-ray pulsar, AX~J1841.0$-$0536,
serendipitously found in the Scutum arm region with the ASCA in two separate 
observations.  The X-ray flux is very faint at the beginning, but exhibits
two flares in the second observation. The flare flux increases by a factor
10 within only $\sim$1 hr.
Coherent pulsations with a period of 4.7394$\pm$0.0008~s 
were detected in the brightest flare phase.
The X-ray spectra in the quiescent and flare phases were fitted
with an absorbed power-law model with a photon index $\sim 1$
plus a narrow Gaussian line at the  center energy of 6.4~keV.
The interstellar column density of $\rm \sim 3\times 10^{22}\ cm^{-2}$
may indicate that AX~J1841.0$-$0536 is located at a tangential point of 
the Scutum arm at $\sim 10$~kpc distance. 
The coherent pulsations, large flux variability and the spectral shape
suggest that AX~J1841.0$-$0536 is a Be/X-ray binary pulsar.

\end{abstract}

\section{Introduction}

The Scutum region near $l\sim \timeform{30D}$ is in the tangential line 
of sight of a galactic arm (the Scutum arm),
and hence shows intensity excesses in the infrared, 
radio and $\gamma$-ray regions (e.g., \cite{hayakawa})
as well as diffuse X-rays (\cite{yamauchi1993}; \cite{sugizaki}).
This region is also rich in X-ray binary sources; 
many transient X-ray sources have also been discovered
(e.g., \cite{koyama}; \cite{yamauchi1995}; \cite{terada}).
From the spectral and flux variability, these X-ray transients are
suspected to be
either neutron-star binaries with high mass or low-mass
stars, or black-hole binaries (\cite{tanaka}). Coherent X-ray 
pulsations have been found from some of them, suggesting Be/X-ray binaries. 

Among them, an unusual source is AX~1845.0$-$0433, which
exhibited violent flares with a very rapid rise time of $\lesssim 10^4$~s 
(\cite{yamauchi1995}).  Apart from the unusual flares, other characteristics 
are similar to those of high-mass X-ray binaries.
In fact, \citet{coe} reported an O9.5 star as the optical counterpart of
AX~1845.0$-$0433.
Nevertheless, no coherent pulsation has been found.

In the ASCA survey and pointing observations on the Scutum arm region,
we found another unusual source, AX~J1841.0$-$0536,  which also exhibited  
violent and rapid flares \citep{bamba1999}.
This paper reports on a detailed analysis of this new transient X-ray source, 
particularly on the detection of coherent
pulsations and phase resolved spectral analyses.

\section{Observations and Data Reduction}

The first ASCA observation of a region around
$(l, b) = (\timeform{26.9D}, -\timeform{0.1D})$ was made on
1994 April 12, or MJD 49454.677--49454.701,
as a part of the Scutum arm survey project (hereinafter, obs.\ 1).
A longer exposure follow-up observation was made  
on 1999 October 3--4, or on MJD 51454.252--51455.463 (obs.\ 2).
ASCA carried four XRTs (X-Ray Telescopes, \cite{serlemitsos})
with two GISs (Gas Imaging Spectrometers, \cite{ohashi}) and 
two SISs (Solid-state Imaging Spectrometers, \cite{burke}) on the focal planes.
Since our target was out of the SIS field in both observations,
we do not refer to the SIS in this paper.
The GISs were operated in the nominal PH mode
with a time resolution of 62.5~ms (High bit-rate) or 0.5~s (Medium bit-rate).

We rejected the GIS data obtained in the South Atlantic Anomaly,
in low cut-off rigidity regions ($<6$~GV), 
or when the target's elevation angle was low ($< 5^\circ$).
Particle events were removed by the rise-time discrimination method
\citep{ohashi}.
After these screenings, the total available exposure times of obs.\ 1 and 2
were $\sim 2$~ks and $\sim 36$~ks, respectively.
To increase the statistics, the data of the two detectors, GIS-2 and GIS-3,
were combined in a following study.

\section{Results}

\subsection{X-Ray Image}

\begin{figure}[hbtp]
 \begin{center}
  \FigureFile(70mm,70mm){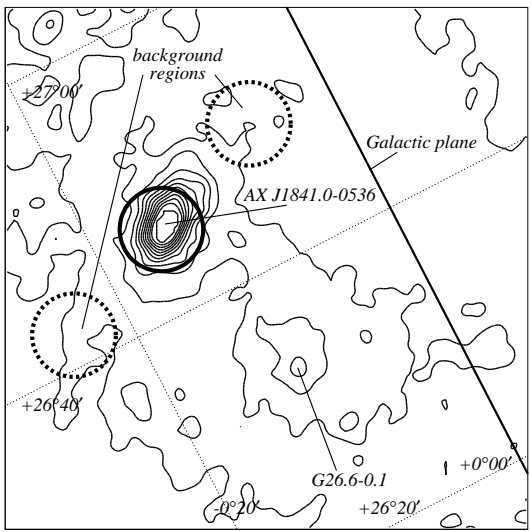}
  \caption{X-ray image around AX~J1849.8$-$0530 in the 2.0--7.0~keV band
	combined with both observations,
	with the galactic coordinates.
	G26.6$-$0.1 is an unidentified hard diffuse source
	(A. Bamba et al. unpublished).
	The source and background regions for the spectral and timing analyses
	are also indicated by the solid and dotted circles, respectively.}
  \label{figure1}
 \end{center}
\end{figure}

A transient source is found  in both obs.\ 1 and obs.\ 2.
Figure~\ref{figure1} shows a point-like source image
located near the galactic plane.
The source position is determined by the method of  
\citet{gotthelf2000} and is  
(R.A., Dec.)$_{\rm J2000}$ = (\timeform{18h41m50.2s}, $-$\timeform{05D36'00''})
with an uncertainty of $\sim 1^\prime$.
Therefore, we designate this source as AX~J1841.0$-$0536
(this source was at first named AX~J1841.0$-$0535.8 based on
less-accurate positioning; \cite{bamba1999}).
No X-ray source has been found within the error region in the SIMBAD database.\footnote{see {\tt http://simbad.u-strasbg.fr/sim-fid.pl}.}
However, this source is in an error region
$(\timeform{0.4D} \times \timeform{4D})$ 
of a previous X-ray transient (the Ginga source No.\ 2 in \cite{koyama}).

\subsection{Light Curve and Timing Analyses}

\begin{figure}[hbtp]
 \begin{center}
  \FigureFile(90mm,150mm){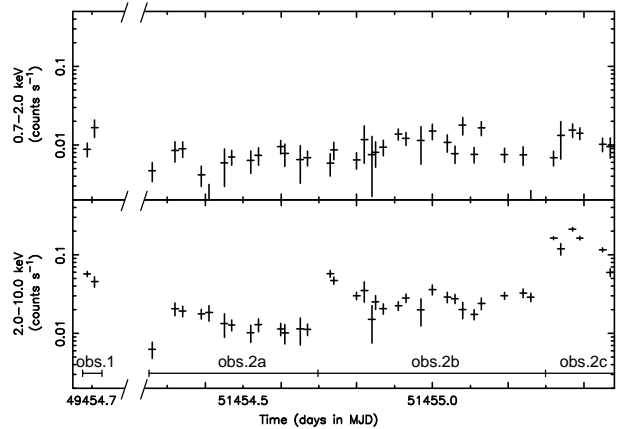}
  \caption{X-ray light curves of AX~J1841.0$-$0536
	observed with GIS-2 and GIS-3 (bin time = 1600~s)
	in obs.~1 and obs.~2.
	The upper and lower panels are those  in the 0.7--2.0~keV
	and 2.0--10.0~keV bands, respectively.
	For the spectral analyses (see text), obs.~2 was divided
	into three epochs (obs.~2a, obs.~2b, and obs.~2c),
	as indicated in the lower panel.}
  \label{figure2}
 \end{center}
\end{figure}

The light curves of AX~J1841.0$-$0536 in both observations
are shown in figure~\ref{figure2}.
The X-ray flux is highly variable with
two flares of the on-set epochs at MJD 51454.7 and MJD 51455.3
with an interval of only  $\sim 0.6$~d.
As can be seen in figure~\ref{figure2}, the flares are found in the hard 
bands, while the soft band fluxes are  essentially constant.
A remarkable fact is that
the on-set time of both the flares are extremely rapid;
the fluxes increased by one order of magnitude within 1~hr.

\begin{figure}[hbtp]
 \begin{center}
  \FigureFile(70mm,50mm){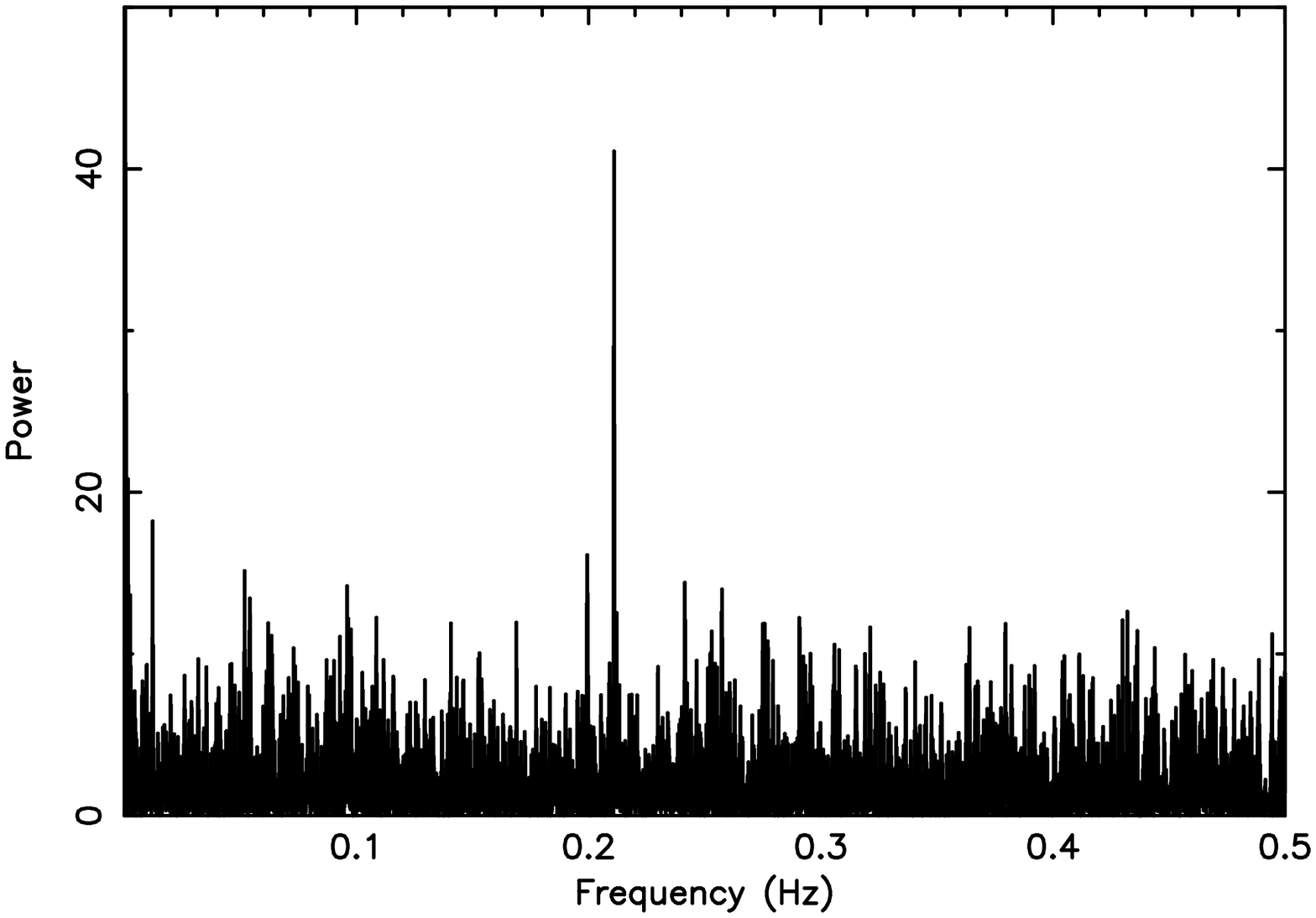}
  \FigureFile(70mm,50mm){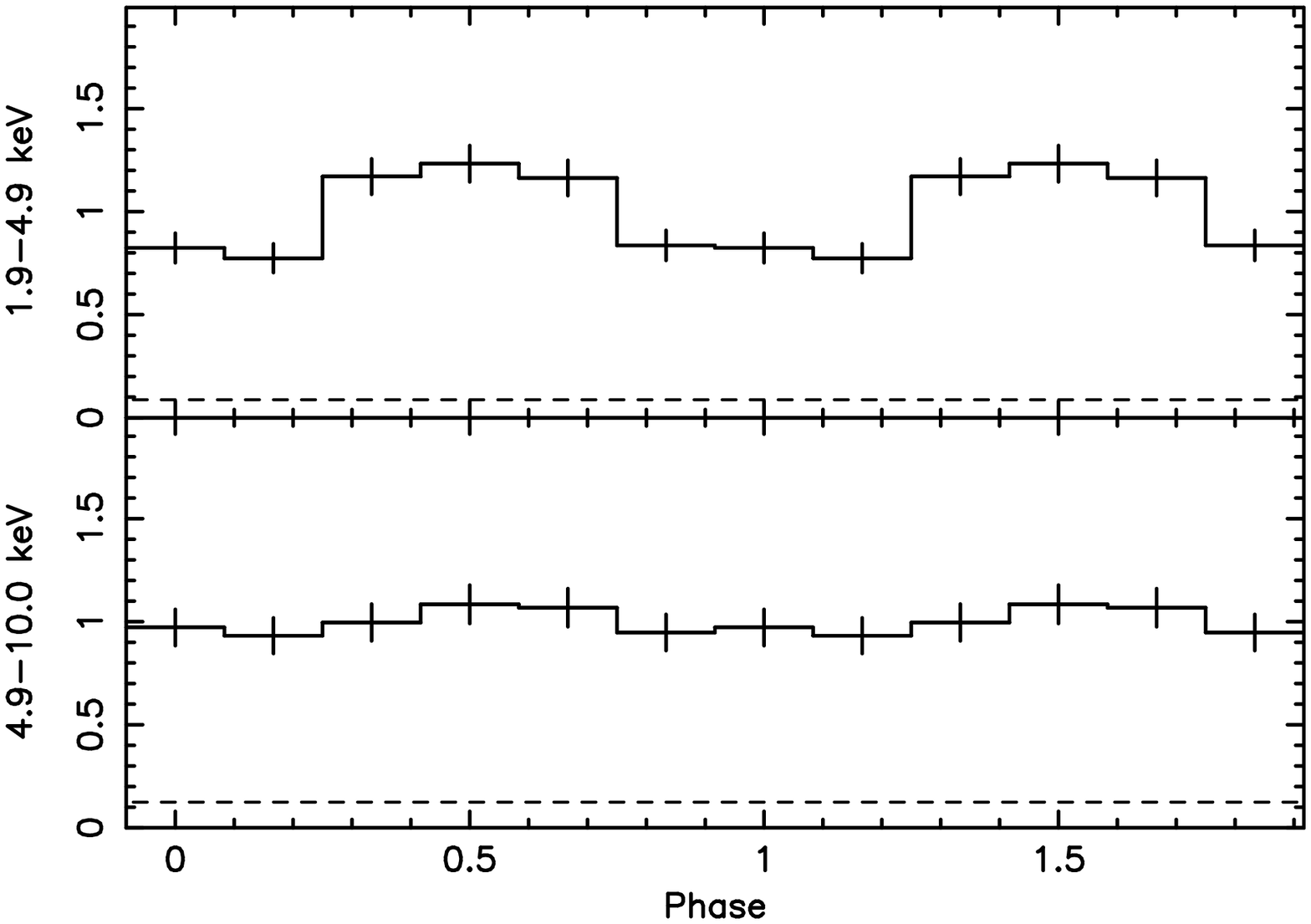}
  \caption{Upper panel: Power-density spectrum in obs.\ 2c.
	A significant peak is detected around $\sim 0.211$~Hz.
	Lower panel: Pulse profiles in the 1.9--4.9~keV (upper panel)
	and the 4.9--10.0~keV (lower panel) bands in obs.\ 2c.
	The vertical axes indicate the normalized intensity per bin. 
	The horizontal axis indicates the pulsation phase
	and the epoch of the beginning of the phase corresponds
	to MJD 51455.300.
	The background levels are indicated by the broken lines.}
  \label{figure3}
 \end{center}
\end{figure}

To study the time evolution, we divided obs.\ 2 into three epochs:
obs.\ 2a (MJD 51454.252--51454.700), obs.\ 2b (MJD 51454.700--51455.300),
and obs.\ 2c (MJD 51455.300--51455.463),
as shown in figure~\ref{figure2}.
We then performed detailed timing analyses of the brightest flare episode
(obs.\ 2c).
We extracted $\sim 1000$~photons, 
including $\sim 90$ background photons, 
in the 1.9--4.9~keV band from a 3$^\prime$ radius circle around the source
(the solid circle in figure 1), corrected the photon arrival times to
those at the barycenter and applied an FFT (Fast Fourier Transform) algorithm.
The resulting power-density spectrum,
figure~\ref{figure3} (left), shows a peak at $\sim$ 0.211~Hz. 
With 13471 frequencies examined, the chance probability of
producing such a large peak is $\sim 9.8\times 10^{-6}$, and hence the 0.211~Hz
oscillation is highly significant.
We next  performed an epoch-folding search and determined
a more accurate pulse period of $P = 4.7394\pm 0.0008$~s.
Figure~\ref{figure3} (right) shows the folded pulse profile 
in the soft (1.9--4.9~keV) and hard (4.9--10.0~keV) bands. 

The pulse shapes are sinusoidal in both energy bands,
but the background-subtracted pulse fraction is larger in the soft (28\%) band
than in the hard (8\%) band.
Although same analyses, FFT and epoch folding, were performed
for the other data sets, no significant power was found
with the $3\sigma$ upper limit of the background-subtracted pulse fraction
in the soft band of  about  
45\% and 12\%  for obs.\ 2a and \ 2b, respectively.

\subsection{Spectral Analyses}

We made time-sliced X-ray spectra using photons
from the same region of the timing analysis
(in the 3$^\prime$ radius circle around the source; see figure 1).
The contribution of the galactic ridge emission is not negligible
for AX~J1841.0$-$0536 located closely on the galactic plane
at $(l, b) = (\timeform{26.7D}, \timeform{0.1D})$.
Contamination from an  unidentified hard diffuse source G26.6$-$0.1
(A. Bamba et al. unpublished) and vignetting may not be ignored. 
We thus made the background spectrum by adding data from two source-free
regions, indicated by the dotted-circles in figure~\ref{figure1}.
These two regions were selected to compensate for
the latitude-dependent galactic ridge emission \citep{kaneda} and
contamination of G26.6$-$0.1. The vignetting  is also the same
as the source region.

The background-subtracted spectrum in the brightest flare phase 
(the second flare, or obs.\ 2c)
was at first fitted  with a power-law function. The  absorption column was 
estimated from the cross sections of
Morrison and McCammon (1983) with solar abundances.
Although this model is statistically acceptable,
we  see a systematic data excess at around 6--7~keV.
Adding  a narrow Gaussian line to the model reduced the
$\chi^2$ from 51.2/52 to 43.2/50 ($\chi^2$/degree of freedom); thus,
the presence of the line is significant.
The center energy of the line is  $6.40\pm 0.12$~keV
(hereinafter, the errors are of 90\% confidence level,
unless otherwise mentioned),
implying that the line is due to neutral or low-ionized irons.

An absorbed thin-thermal plasma model fitting, on the other hand,
requires  unrealistically high  temperature of $>$ 80~keV,
and is hence unlikely.

\begin{figure}[hbtp]
 \begin{center}
  \FigureFile(70mm,50mm){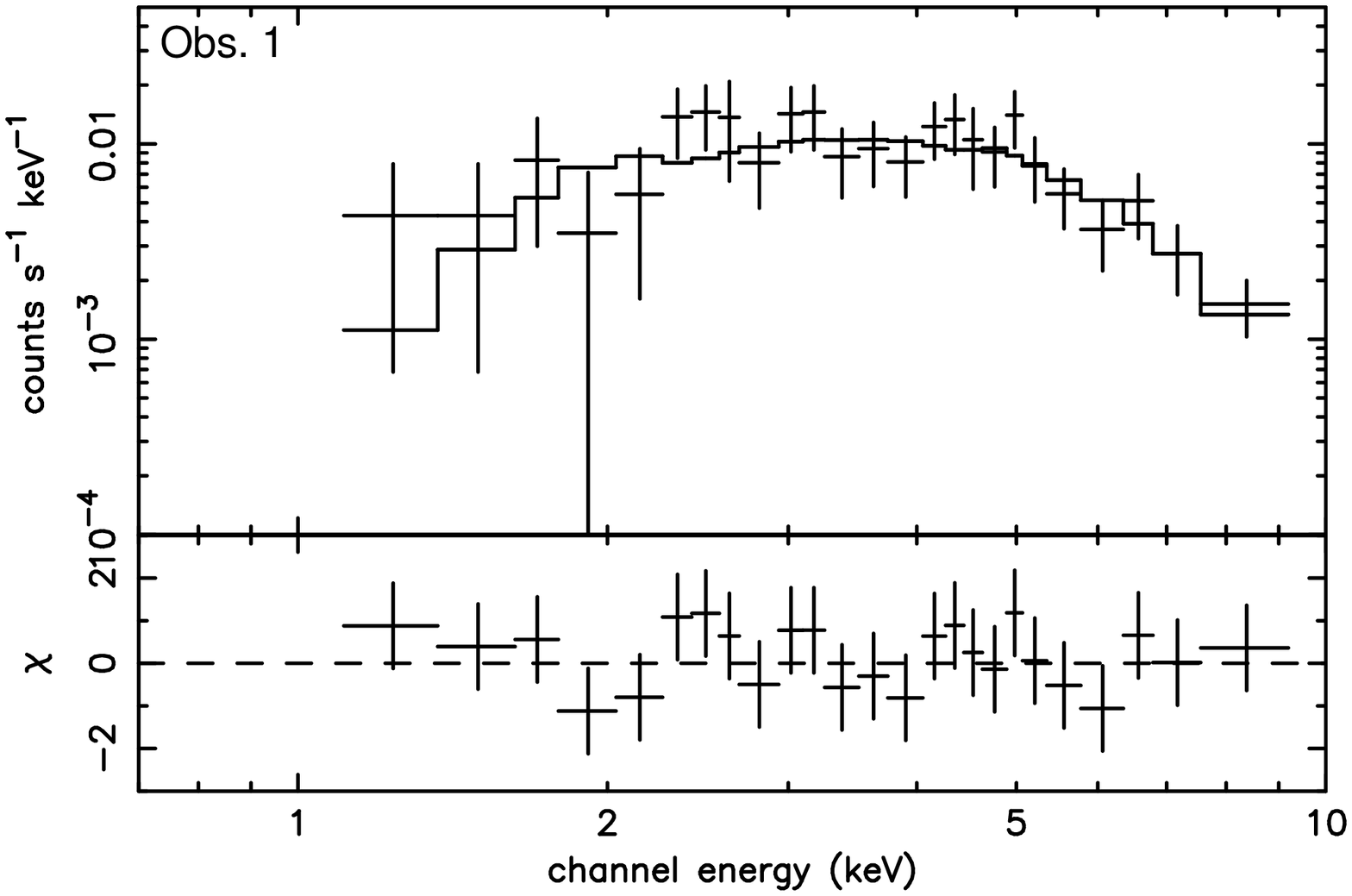}
  \FigureFile(70mm,50mm){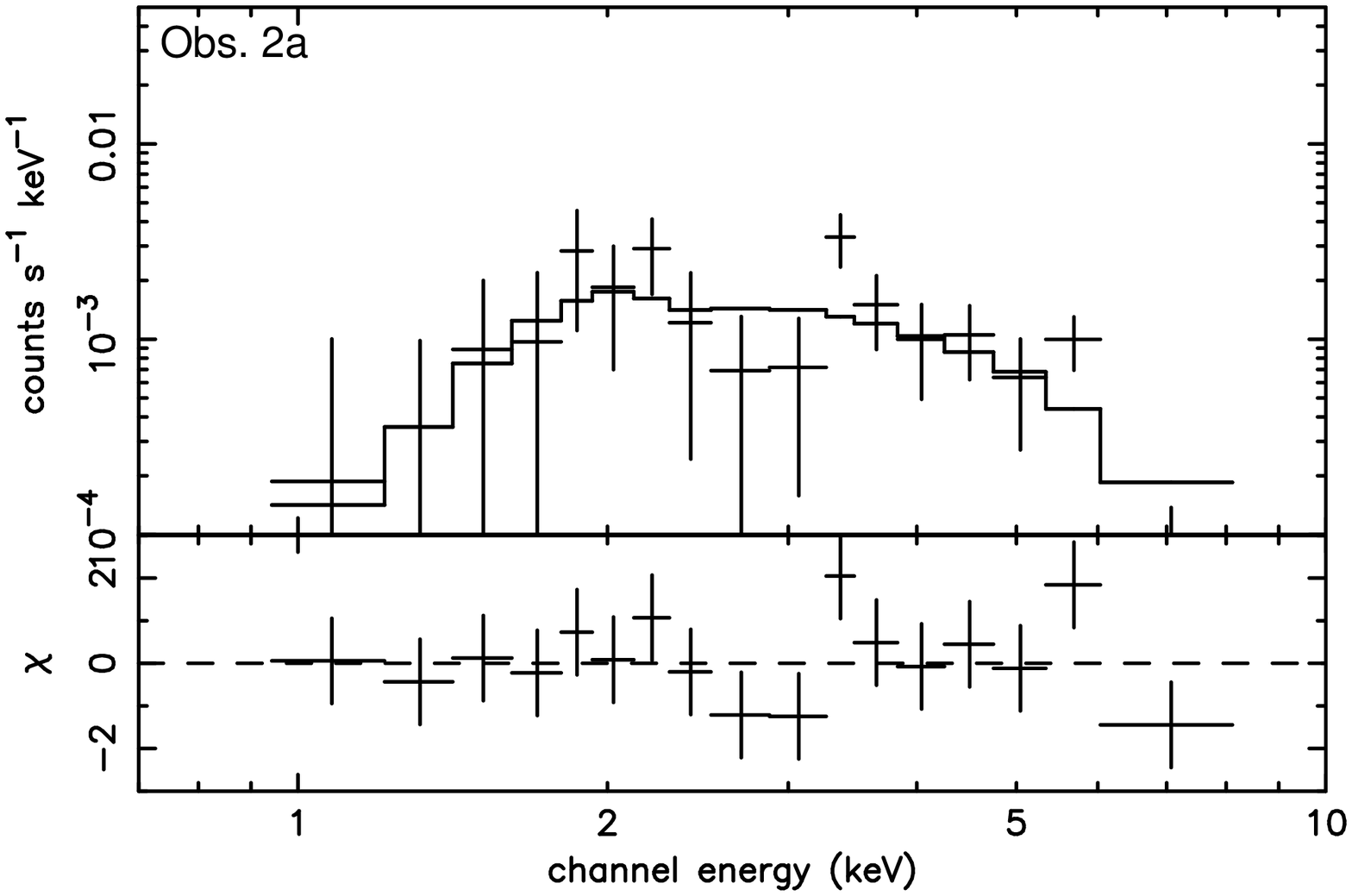}
  \FigureFile(70mm,50mm){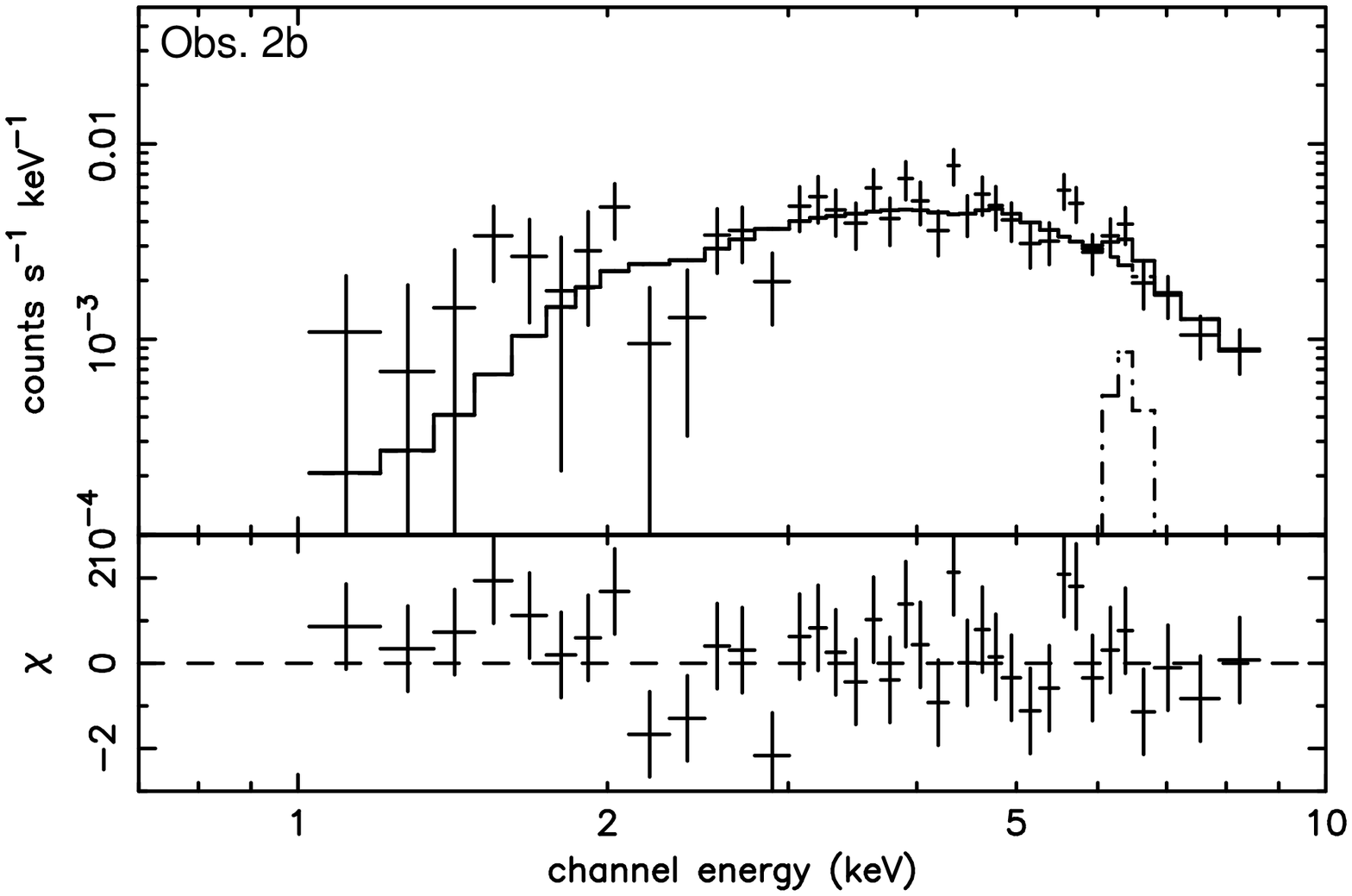}
  \FigureFile(70mm,50mm){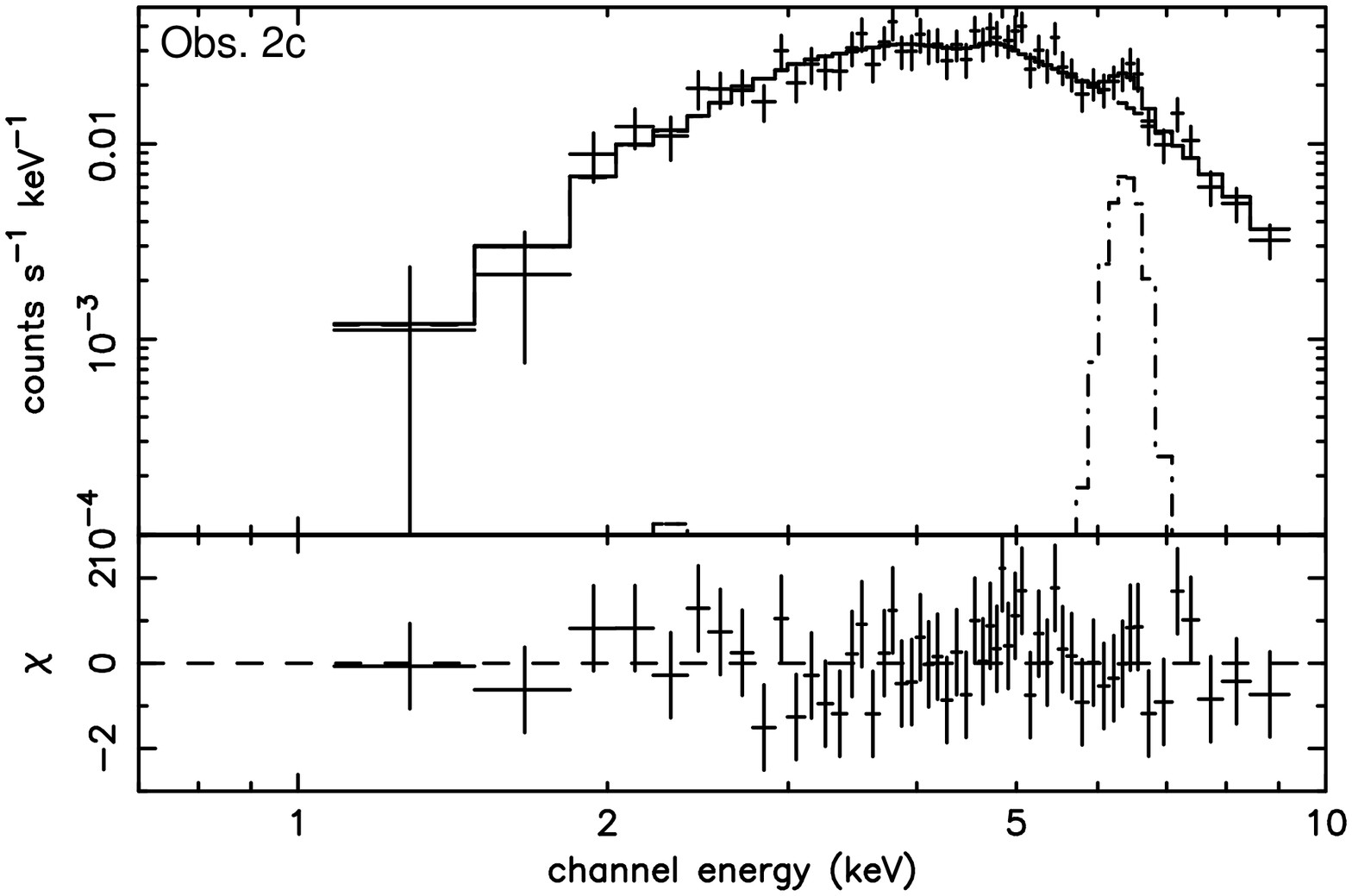}
  \caption{Background-subtracted spectra (crosses) and the best-fit models
	(solid histograms). Data residuals  are shown in the lower panel
	of each figure.}
  \label{figure4}
 \end{center}
\end{figure}

We then fit the background-subtracted spectra of the other epochs 
with a power-law plus a Gaussian model,
fixing the center energy of the line to be 6.40~keV
(the best-fit value in obs.\ 2c). 
These fittings are acceptable for all epochs.
The spectra and the best-fit models for the
four separate epochs are shown in figure~\ref{figure4}, while
the best-fit parameters are listed in table~\ref{table1}.

\begin{longtable}{p{10pc}cccc}
  \caption{Best-fit parameters for AX~J1841.0$-$0536
	for an absorbed power-law and a Gaussian model.$^{\ast}$}
  \label{table1}
\hline\hline
 Parameters & obs.\ 1 & obs.\ 2a & obs.\ 2b & obs.\ 2c
\endfirsthead
\endhead
\hline
Power-law & & & & \\
\ \  $\Gamma$\dotfill & 1.0 (0.3--1.9) & 2.0 (0.9--3.6) & 0.7 (0.1--1.3) & 1.1 (0.9--1.4) \\
Gaussian & & & & \\
\ \ Center (keV)\dotfill & 6.4 (fixed) & 6.4 (fixed) & 6.4 (fixed) &6.4 (6.3--6.5) \\
\ \ EW$^\dagger$ ($\times 10^2$~eV)\dotfill & ($<$ 4.7) & not determined & 1.9 ($<$ 4.3) & 2.3 (1.0--3.8)\\
$N_{\rm H}^\ddagger\ (\times 10^{22}{\rm cm}^{-2})$\dotfill & 3.2 (1.1--6.4) & 3.2 (0.8--8.4) & 4.3 (2.0--7.9) & 7.2 (6.0--8.6) \\ \hline
Flux$^\S$ (erg cm$^{-2}$s$^{-1}$)\dotfill & $2.0\times 10^{-11}$ & $1.9\times 10^{-12}$ & $1.3\times 10^{-11}$ & $9.5\times 10^{-11}$ \\
Luminosity$^{\S \amalg}$\ $\rm (erg\ s^{-1})$\dotfill & $2.3\times 10^{35}$ & $2.3\times 10^{34}$ & $1.5\times 10^{35}$ & $1.1\times 10^{36}$ \\
$\chi^2$/d.o.f$^\#$\dotfill & 13.2/21 & 15.1/13 & 41.9/34 & 43.2/50\\ \hline\hline
\multicolumn{5}{l}
{$^{\ast}$ Errors and upper limits are at 90\% confidence
	for one relevant parameter.}\\
\multicolumn{5}{l}
{$^{\dagger}$ Equivalent width.}\\
\multicolumn{5}{l}
{$^\ddagger$ Assuming the solar abundance ratio \citep{morrison}.}\\
\multicolumn{5}{l}
{$^\S$ Estimated in the 2.0--10.0~keV band.}\\
\multicolumn{5}{l}
{$^{\amalg}$ Assumed distance is 10~kpc (see section \ref{section4_1}.)}\\
\multicolumn{5}{l}
{$^\#$ Degree of freedom.}\\
\end{longtable}

\section{Discussion}

% \subsection{Distance and Luminosity}
\label{section4_1}

AX~J1841.0$-$0536 is in the error region
$(l = \timeform{26.6D}\pm\timeform{0.2D},
b = -\timeform{0.5D}\pm\timeform{2.0D})$
of a previous X-ray transient No.~2 in \citet{koyama}. 
Although the position uncertainty is large, the X-ray features, 
power-law index of 1.5$\pm$0.3, flux of
$\rm 2\times 10^{-11}\ erg\ cm^{-2}\ s^{-1}$ and
log ($N_{\rm H}$) of 22.7$\pm$ 0.3
are similar to those of  AX~J1841.0$-$0536 in obs.\ 1 and obs.\ 2b.
Therefore these two transients are likely to be the same source.
From the flux history, we suspect that
the quiescent level of AX~J1841.0$-$0536 is 
on the order of $\rm 2\times 10^{-12}\ erg\ cm^{-2} \ s^{-1}$,
and that flares may occur very frequently.
More bright flares of a level of $\rm 10^{-10}\ erg\ cm^{-2}\ s^{-1}$ are,
however not frequent.

Although the limited statistics allows a constant-value hypothesis of
$N_{\rm H}$, we see a hint that $N_{\rm H}$ increase with increasing flux,
which suggests that the circum-stellar absorption becomes larger in the flare,
due to the larger gas accretion.
Therefore, the interstellar absorption should be regarded as
the minimum value of $\sim 3.2_{-2.4}^{+6.2}\times 10^{22}\ {\rm cm\ s}^{-1}$
(obs.\ 1 and obs.\ 2a). The large error of  $N_{\rm H}$ does not allow us 
to reliably estimate the source distance. Nevertheless, we simply estimate
the distance to AX~J1841.0$-$0536 to be 1-10~kpc, assuming
the average density in the galactic plane to be 1 H cm$^{-3}$.
This value is consistent with  that the  transient source
is located in the tangential point of the Scutum arm.

\citet{koyama} reported many X-ray transient sources in the Scutum arm
with a flare of the order of $\rm 10^{36}\ erg\ s^{-1}$, and suggested that
most are  Be/X-ray binaries.
Since the hard X-ray spectrum, neutral iron line, erratic flux variability 
and coherent pulsations are characteristic of Be/X-ray binaries,  
AX~J1841.0$-$0536 would be a member of this sub-class of X-ray pulsars.
No optical counterpart has been found to confirm this classification.
The luminosity is consistent with that of a
``Type I outburst'' of Be/X-ray binaries \citep{negueruela}.

From a semi-empirical relation
of the orbital period vs. pulse period and the X-ray luminosity
(\cite{corbet1984}; \cite{corbet1999}),
we suggest that the orbital period of AX~J1841.0$-$0536 is $\sim $25~days.  
Be/X-ray binaries often exhibit flares with an interval of the orbital period 
(c.f. 4U~1223$-$62, \cite{watson}; V~0332+53, \cite{stella})
due to a gas surge during passage near to the peri-astron or across the
equatorial plane.
From AX~J1841.0$-$0536, we observed two subsequent flares separated by
$\rm \sim 0.6\ d$,
which is much shorter than not only the predicted orbital period,
but also the shortest orbital period known for a Be/X-ray binary pulsar, 
16.7~days for A0538$-$66 \citep{corbet1997}.

\citet{parmar} reported that
quasi-periodic flares with a 3.96 hr interval in the limited episode
of a Be/X-ray binary  EXO~2030+375, which has spin and orbital periods of
$\sim 42$~s and 46.0~day respectively \citep{stollberg},
are both different from the flare interval.
They inferred that the quasi-periodic rapid flares, 
fast rise and exponential decay, are due to instabilities
in the either magnetosphere (e.g., \cite{lamb}) or the accretion disk
(e.g., \cite{taam1984}).
Asymmetric flow of a stellar wind (e.g., \cite{taam1988}),
or non-homogeneity of the OB star wind (e.g., \cite{baade}) may also be
a possible scenario.
The same process can be applied to the two flares of AX~J1841.0$-$0536,
although the detailed structure, the flare profile, number, size and so on
are significantly different from those of EXO~2030+375.

A remarkable feature of AX~J1841.0$-$0536 is an eruptive on-set of a 
flare with a rise-time of $\sim$1~hour.
We note that other transient sources, AX~J1845.0$-$0433 \citep{yamauchi1995}
and XTE~J1739$-$302 (\cite{smith}; \cite{sakano}),
exhibit multiple flares with a fast-rise of a few hours,
although no coherent pulsation has been found.
We suspect that
these new transients comprise a new sub-class of X-ray binaries,
which provides a crucial test for the accretion-flow instability.
Therefore, further monitor observations
on these unusual X-ray binaries are encouraged. 
 
\vspace{1cm}

We thank all of the ASCA galactic ridge survey members.
This research made use of the SIMBAD database,
operated at CDS, Strasbourg, France.
We also thank an anonymous referee, D. Yonetoku and K. Imanishi for their useful comments and suggestions.
A.B. and J.Y. are supported by JSPS Research Fellowship for Young Scientists.

\end{document}